\documentclass[iop]{emulateapj}

\newcommand{\Msun}{M\ensuremath{_{\odot}}~}

\shorttitle{Mass Loss via Stellar CMEs}
\shortauthors{Aarnio, Matt, and Stassun}

\begin{document}

\title{Mass Loss in Pre-Main Sequence Stars via Coronal Mass Ejections and Implications for Angular Momentum Loss}

\author{Alicia N.\ Aarnio}
\affil{Astronomy Department, University of Michigan, 830 Dennison Building, 500 Church Street, Ann Arbor, MI 48109 USA 
\email{aarnio@umich.edu}}

\author{Sean P.\ Matt}
\affil{Laboratoire AIM Paris-Saclay, CEA/Irfu Universit\'e Paris-Diderot CNRS/INSU, 91191 Gif-sur-Yvette, France}

\author{Keivan G.\ Stassun}
\affil{Department of Physics and Astronomy, Vanderbilt University, Nashville, TN 37235 USA} 
\affil{Department of Physics, Fisk University, Nashville, TN 37208 USA}

\begin{abstract}
We develop an empirical model to estimate mass-loss rates via coronal mass ejections (CMEs) for solar-type pre-main-sequence (PMS) 
stars. Our method estimates the CME mass-loss rate from the observed energies of PMS X-ray flares, using our empirically determined 
relationship between solar X-ray flare energy and CME mass: 
$\log(M_{\rm CME}{\rm~[g]}) = 0.63 \times \log(E_{\rm flare} {\rm [erg]}) -2.57$.
Using masses determined for the largest flaring magnetic structures observed on PMS stars, we suggest that this solar-calibrated 
relationship may hold over 10 orders of magnitude in flare energy and 7 orders of magnitude in CME mass. The total CME mass-loss 
rate we calculate for typical solar-type PMS stars is in the range $10^{-12}$--$10^{-9}$ \Msun yr$^{-1}$. We then use these CME 
mass-loss rate estimates to infer the attendant angular momentum loss leading up to the main sequence. Assuming the CME outflow rate 
for a typical $\sim$1~\Msun T Tauri star is $< 10^{-10}$ \Msun yr$^{-1}$, the resulting spin-down torque is too small during the 
first $\sim$1~Myr to counteract the stellar spin-up due to contraction and accretion. However, if the CME mass-loss rate is 
$\ga 10^{-10}$ \Msun yr$^{-1}$, as permitted by our calculations, the CME spin-down torque may influence the stellar spin evolution 
after an age of a few Myr.
\end{abstract}

\keywords{Stars:activity --- Stars:evolution --- Stars: mass-loss --- Stars: pre-main sequence --- Stars: solar-type}

\section{Introduction\label{sec_intro}}

Solar-type pre--main-sequence (PMS) stars typically evince magnetic activity in a variety of forms, including strong and time-variable 
coronal emission at X-ray wavelengths \citep[e.g.][]{guedel08,Preibisch:2005}, strong and time-variable chromospheric emission at 
ultraviolet wavelengths \citep[e.g.][]{yang12,petrov11,valenti93} and in chromospheric emission lines 
\citep[such as H$\alpha$; e.g.][]{white03,fernandez95,walter81}, large spot-modulated photometric variability 
\citep[e.g.][]{herbst07,stassun99,herbst94}, and non-thermal polarized radio continuum emission is a classic indicator of magnetic 
activity \citep[e.g., the PMS star Hubble 4,][]{Skinner:1993}. These observations have motivated direct Zeeman measurements revealing 
field strengths of typically a few kG \citep[e.g.][]{johnskrull07,reiners07,johnskrull04}. The strong magnetic fields associated with 
this ubiquitous activity are thought to be central to a number of key physical processes in PMS stars, including specifically the 
transfer of mass and of angular momentum from and to the circumstellar environment. 

X-ray activity observed in PMS stars bears a striking resemblance to solar X-ray activity, albeit scaled up by several orders of magnitude 
in both energy and frequency of occurrence \citep[e.g.,][]{Peres:2004,Getman:2005,Peres:2006,Audard:2007}. Generally, PMS X-ray flares are 
well described by standard solar flare models \citep[e.g.,][]{Reale:1997}, though in the case of some exceptionally long-lived
flares it has been necessary to add self-eclipse of loop emission by the star \citep[i.e.,][]{Skinner:1997,Johnstone:2012}. 
In addition, with the exception of a few very strongly accreting systems where the X-ray
variability is at least in part accretion driven \citep[e.g., V1647 Ori;][]{kastner04,hamaguchi12},
large-scale studies of simultaneous X-ray and optical variability in PMS stars show that the X-ray variability is principally caused by 
coronal activity \citep{stassun06,stassun07}. Young, solar-type stars appear, then, to have flaring magnetic field configurations that 
behave like, and can be interpreted as, scaled-up solar type flaring fields \citep[see also][for a recent overview]{feigelson07}. 

During the early, active accretion phase of a PMS star's evolution \citep[first few Myr; e.g.][]{haisch01}, the strong stellar field may 
be important in governing the interaction of the star with its protoplanetary disk. This magnetic star-disk interaction likely controls the 
funneling of circumstellar material from the disk onto the star \citep[e.g.][]{shu94}, is likely central to the launching of outflows 
including jets and winds \citep[e.g.][]{Hayashi:1996,Matt:2005}, and therefore is likely a key mediator in the net flux of mass and angular 
momentum onto and away from the young star \citep[e.g.][]{Matt:2005,Matt:2010,Orlando:2011}.

Even after the phase of active accretion subsides, scaled-up solar-type winds and elevated solar-type X-ray activity do not decline to 
present-day solar levels until approximately 1 Gyr \citep{Wood:2004,Wood:2005,Guedel:1997}, potentially affecting the 
ongoing evolution of the star's angular momentum and the circumstellar environments of young planets \citep[e.g.][]{vidotto10}. 

Extremely large magnetic loops---extending out to tens of stellar radii from the stellar surface---have been inferred among the most powerful 
X-ray flaring PMS stars from the {\it Chandra} Orion Ultra-deep Project (COUP) sample \citep[e.g.][]{Favata:2005,Massi:2008,Skelly:2008}, 
implying very large magnetic lever arms that are potentially able to shed significant angular momentum \citep{Aarnio:2009,Aarnio:2012}. 
While it was at first appealing to picture these large magnetic loops as being linked to circumstellar disks, perhaps either arising from a 
star-disk reconnection event \citep{Orlando:2011} or simply maintaining loop stability via anchoring to the disk, \citet{Aarnio:2010} showed 
that the majority of these stars lack massive disks within reach of the loops. Indeed, earlier work by \citet{Getman:2008} came to similar 
conclusions, inferring from the apparent absence of large magnetic loops among stars with close-in disks that perhaps it is the absence of the 
truncation by an inner dust disk which allows these loops to be so large in the first place. 

On the Sun, the eruption of such magnetic structures can drive coronal mass ejections (CMEs), which are observed in Thomson-scattered optical 
light \citep[cf., ][]{Vourlidas:2006}. While the faint scattered light of CMEs cannot yet be directly detected from most other stars, there 
have been recent observations of young stars with the FUSE satellite \citep{Leitzinger:2011} and the detection of a large flare-associated mass 
ejection in the PMS star Z~CMa \citep{Stelzer:2009,Whelan:2010}. More generally, CMEs are expected from low-mass stars by analogy to the Sun as well as from 
basic escape velocity considerations for the flaring material; indeed, CMEs are believed to have been detected on M and K dwarfs from balmer 
line asymmetries, transient spectral line absorption components, and EUV dimmings \citep[cf., ][and references therein]{Leitzinger:2011}.
In particular, the extreme X-ray flares observed on PMS stars are expected to have associated extreme CMEs \citep[e.g.,][]{Aarnio:2011a}, 
the ramifications of which---for the evolution of the star, for the protoplanetary environment, and for the star's angular momentum 
evolution---depend on the resulting mass-loss rate and the evolution of that mass loss over the course of the star's PMS evolution. 

A number of studies have used emission line tracers to empirically estimate the mass-loss rates from PMS winds during the actively accreting, 
classical T~Tauri star (CTTS) phase \citep[see, e.g.,][and references therein]{kwan07,edwards06,edwards03}, and such accretion-powered PMS winds 
have been explored in detail theoretically \citep[e.g.][]{Matt:2005,Matt:2008b,Cranmer:2008,Matt:2008c,Cranmer:2009,Matt:2010}. Meanwhile, 
reliable empirical measures of mass-loss rates for non-accreting PMS stars, particularly from impulsive CME-type events, are virtually nonexistent, 
and simulations of winds from non-accreting, weak-lined T~Tauri stars (WTTSs) have only recently been investigated \citep{vidotto09a,vidotto09b,vidotto10}. 

Here we aim to make first empirical estimates of mass-loss rates for PMS stars resulting from CMEs, and then use these mass-loss rate estimates 
to infer the attendant angular momentum loss leading up to the main sequence. Specifically, we develop a method to estimate CME mass-loss rates 
from observed PMS X-ray flares, calibrated to solar observations and models, and then calculate the resulting braking torques in the context of 
PMS stellar evolutionary models. 

We begin by summarizing our previous work to establish an empirical relationship between X-ray flare flux and CME mass for the Sun, and describe 
how we then extend that relationship to solar-type PMS stars (Sec.~\ref{sec_solarecap}). In Section \ref{sec_massloss}, we use our empirical flare/CME 
relationship to derive CME mass-loss rates for PMS stars both empirically (using directly observed flare properties) and analytically (using functional 
representations of flare properties), and furthermore test our methodology on the Sun, verifying that we are able to recover its well measured CME 
mass-loss rate from its observed X-ray flares. Next, in Section \ref{sec_angmom} we calculate the attendant torque on PMS stars exerted by the CMEs, and 
assess the likelihood that these braking torques may serve to prevent the stars from spinning up as they contract toward the main sequence. Finally, 
we discuss our findings in Section \ref{sec_discus} and summarize our conclusions in Section \ref{sec_conclusion}.


\section{A solar-calibrated relationship between PMS X-ray flares and CME masses}\label{sec_solarecap}

In \citet{Aarnio:2011a}, we cross-matched 10 years of solar flare and CME observations in order to determine, for associated flares and CMEs, 
whether there are correlations between flare and CME properties. Approximately 1000 solar flares and CMEs were found to be spatially and temporally 
associated. The resulting relationship between flare flux and CME mass is well fit by a single power law of the form:
\begin{equation}\label{eq_fluxmass}
{\rm log(M} {\rm~[g])} = 18.7 + 0.7 \times {\rm log(F} {\rm~[W~m}^{-2}{\rm ])},
\end{equation}
where M is the CME mass and F the flare flux. Flare flux and CME mass hold this log-linear relationship over a few dex in flux and mass. 

\begin{figure*}[ht]
\centerline{\includegraphics[scale=0.5,angle=90]{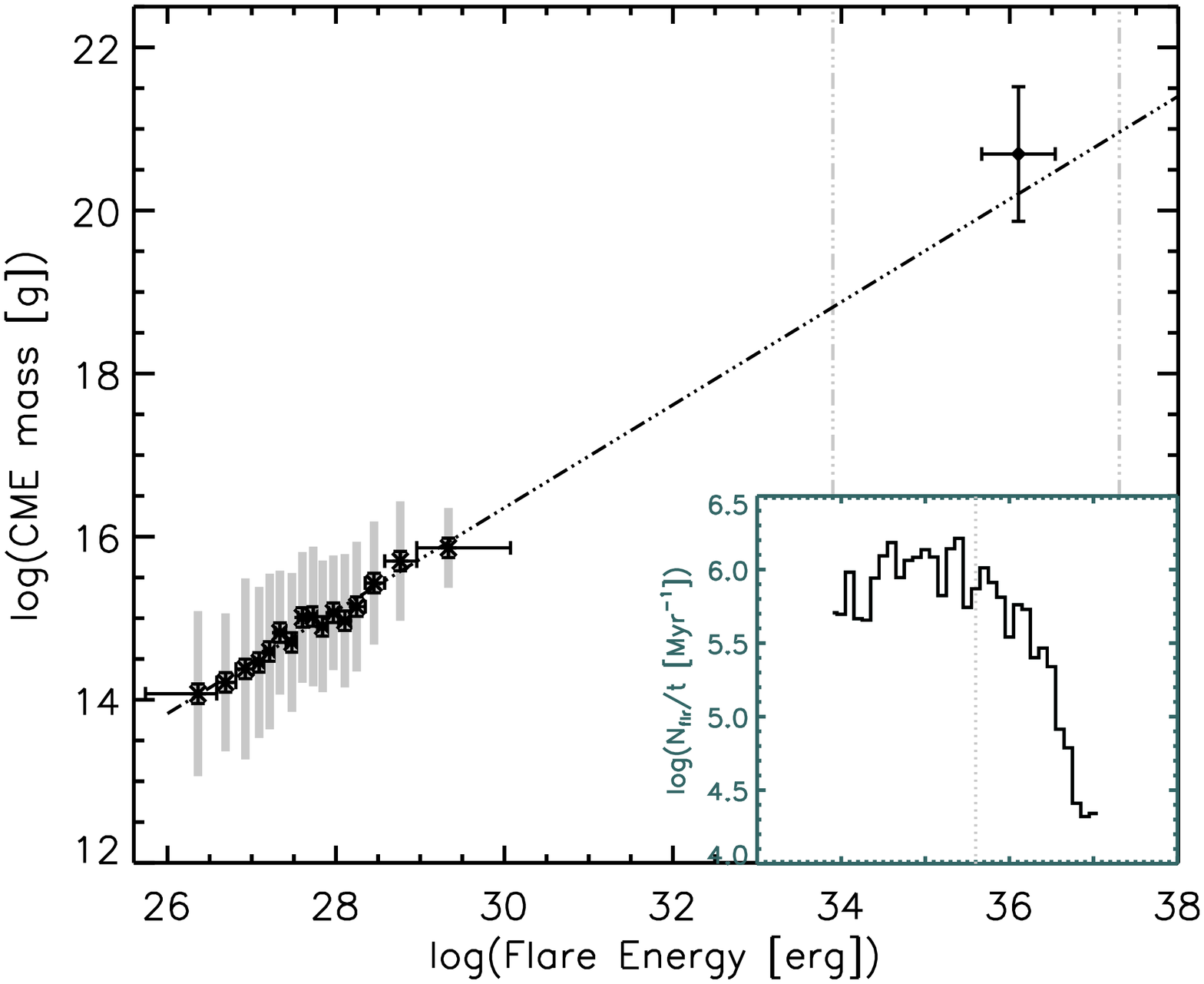}}
\caption{Solar relationship between flare energy and CME mass, transformed from the flux/mass relationship of \citet{Aarnio:2011a}. 
The line is a fit to the solar data over the energy range of observed solar flares, $10^{26}$--$10^{30}$ erg, and then extrapolated to 
the regime of T Tauri star flare energies (represented by vertical lines). The single point at upper right shows the mean CME mass and 
flare energy for the 32 T Tauri star mega-flares observed by COUP \citep{Favata:2005}, and the error bars represent the standard 
deviations of those mean values. The extrapolated solar relationship fits this point well. {\it (Inset:)} The energy distribution 
of flare rates for T Tauri stars observed by COUP \citep{Colombo:2007}. The vertical dotted line at $10^{35.6}$ erg represents the energy 
above which the observed flare sample is complete. } 
\label{fig_massenergy}
\end{figure*}

Solar flares are, by convention, classified by their peak X-ray flux, while for stellar X-ray flare observations, flare luminosities are 
reported. Therefore, to relate our solar flare flux/CME mass relationship to observations of PMS stellar flares, we convert the solar flare X-ray 
fluxes into energies, and re-frame the CME mass/flare flux relationship into a CME mass/flare energy relationship. To determine solar flare energy, 
we integrate the flux of the solar X-ray flare light curves from flare start to end time. For simplicity, we approximate the flares by their peak 
energy and assume a linear decay. Specifically, each flare's total energy output equals one half of the observed peak flux, times the observed flare 
duration, times 4$\pi$(1AU)$^2$. Our final solar CME mass/flare energy relationship is shown in Figure \ref{fig_massenergy}, and it has the functional 
form:
\begin{equation}\label{eq_energymass}
M = K_{\rm M}E^{\beta};
\end{equation}
in this and all following cases, M will denote the CME mass and E the flare energy. K$_{\rm M}$ is (2.7$\pm$1.2)$\times$10$^{-3}$ in cgs units, and 
$\beta =$ 0.63$\pm$0.04.

In adopting this relationship, we take advantage of the empirical correlation between associated flares and CMEs, but 
we make no assumption or statement of a causal relationship between flares and CMEs. Indeed, it is well established 
that not every flare is associated with a CME, and not every CME is associated with a flare. \citet{Aarnio:2011a} found that 13\% of CMEs are flare 
associated and $\sim$9\% of flares are CME associated; this lack of correspondence is at least partially driven by observational biases, and any 
correction for over-- or underestimating the number of CMEs associated with flares is likely less than approximately a factor of two. The most 
energetic solar flares, however, are the most likely to be CME-associated, and these CMEs are the most massive
\citep[][ to name but a few studies on the matter]{Andrews:2003,Yashiro:2005,Yashiro:2006,Aarnio:2011a}.
Since young stars' flares are very powerful with respect to solar flares and the most energetic solar flares have the highest rate of CME 
coincidence, we could suppose that a flare/CME association probability is close to unity for TTS flares.

Since we cannot yet directly observe CMEs on young stars, we will use our solar-calibrated relationship (eq.~\ref{eq_energymass}) 
as applied to X-ray flare observations of young stars to determine limits on the mass losses from young stars via solar-analog CMEs. 
However, PMS X-ray flare energies are 5--7 orders of magnitude stronger than solar flares, requiring a large extrapolation of the solar flare/CME 
relationship. Therefore, it is prudent to demonstrate that the likely masses of CMEs associated with the extreme X-ray flare energies
observed on PMS stars are in fact consistent with the solar flare/CME relationship. To do this, we have estimated the masses associated with the 
32 most powerful PMS X-ray flares observed in Orion by COUP \citep{Favata:2005}. Strictly speaking, these masses represent the masses of post-flaring 
loops, which might not be equivalent to the CME masses, but in any case by analogy to the Sun should be related to the CME masses. 

We use the flaring 
loop physical parameters for these 32 extreme PMS flares as derived by \citet{Favata:2005}, who used the solar-calibrated uniform cooling loop (UCL) 
model of \citet{Reale:1997} to infer the flaring loop lengths and their densities, the latter inferred from the flare peak emission measures. We adopt 
the typically assumed cylindrical loop shape and the typically assumed loop aspect ratio\citep[radius/length = 0.1][]{Reale:1997}, permitting the 
flaring loop volumes to be calculated. The loop masses then follow from the volumes and densities. In this way, we obtain flaring loop masses ranging 
from 10$^{19}$~g to 10$^{22}$~g for the \citet{Favata:2005} sample. These 32 flaring loop masses and their associated flare energies are shown in 
Figure \ref{fig_massenergy} as the point with error bars at the far upper right, representing the mean and standard deviation of the sample. The large 
scatter of $\sim$1 dex in the masses is likely due in part to the simplistic assumptions of the UCL model upon which the mass estimates are based. For 
example, the UCL model assumes that a single magnetic loop is involved in the observed X-ray flare, while on the Sun, arcades of smaller loops are 
often involved in a given flaring ``event." \citet{Vaananen:2009} suggest that the height of such loop arcades can differ from the UCL inferred loop 
heights by factors of 2--10, consistent with the scatter in loop masses we have estimated in Figure \ref{fig_massenergy}. In any case, the extreme PMS 
CME masses so inferred are consistent to within a factor of 10 with that predicted by the solar-calibrated flare/CME relationship. Note that we have 
not adjusted the fit from the solar data, but merely extrapolated it, which lends strong support to the use of this same solar-calibrated flare/CME 
relationship over the full range of solar to TTS flare energies.


\section{Mass loss via CMEs}\label{sec_massloss}

Having established the form of the relationship between flare energy and CME mass, in this section we derive total CME mass-loss rates for PMS stars, both 
(1) empirically using the observed distribution of flare energies directly (Sections \ref{sec_solarempirical} and \ref{sec_stellarempirical}), and 
(2) analytically using the observed flares to describe a functional form for the distribution of flare energies (Sections \ref{sec_solaranalytical} 
and \ref{sec_stellaranalytical}). First, however, we apply our methodology to the solar case as a confirmation that our method recovers the observed 
CME mass-loss rate for the Sun. 

\subsection{Solar case}\label{sec_solarmassloss}

We start by demonstrating our method for the solar case as a benchmark.  For the sun, we have a measured CME mass distribution, from which we can 
directly calculate a CME mass loss rate to compare to the mass-loss rates that we infer from the X-ray flares. 

In the LASCO CME database \citep{Gopalswamy:2009}, 13,862 CMEs are reported from 1996 to 2006, spanning almost a full solar activity period 
including minimum and maximum. Of those, 6,733 CMEs have well constrained mass measurements, and an additional 1,395 have masses flagged as highly 
uncertain. We show the full rate distribution of reported CME masses (number of CMEs per year as a function of CME mass), including both 
well-constrained and highly uncertain CME masses, in Figure \ref{fig_solardists} (upper middle panel). We have included here the additional 
halo CMEs (those projected toward the Earth, or which are so wide as to appear as if they were) for which \citet{Aarnio:2011a} estimated masses.

Summing up this entire distribution gives a CME mass loss rate of 7.8$\times$10$^{-16}$ \Msun yr$^{-1}$. Put another way, between 1996 and 2006, the Sun 
shed at least 1.563$\times$10$^{18}$ g yr$^{-1}$ via CMEs. This value represents a lower limit to the mass loss rate via CMEs from the Sun, as just under 
half of the reported CMEs have measured masses and there may be CMEs that escape observation, especially at lower masses. \citet{Aarnio:2011a} found 1,153 
of the CMEs with measured masses to be associated with flares (the masses of these CMEs are also shown in Figure \ref{fig_solardists}, upper middle panel); 
these CMEs were used to derive the relationship between flare energy and CME mass (eq.~\ref{eq_energymass}). Even though only $\sim$15\% of the CMEs 
with measured masses are associated with flares, these flare-associated CMEs constitute 40\% of the total observed CME mass loss rate.

The above directly measured CME mass-loss rate thus provides a benchmark that our flare-inferred 
CME mass-loss rate below should reproduce. To compare the CME mass-loss rate that we will infer from flares below to the observed 
solar case, we will find it convenient to represent the solar CME mass distribution (Fig.~\ref{fig_solardists}, upper middle panel)
as a differential CME mass distribution ($dN/dM$), and this is shown in Figure \ref{fig_solardists}, lower middle panel.

\subsubsection{Empirically determined solar CME mass-loss rate}\label{sec_solarempirical}

\begin{figure*}[ht]
\centerline{\includegraphics[scale=0.5,angle=90]{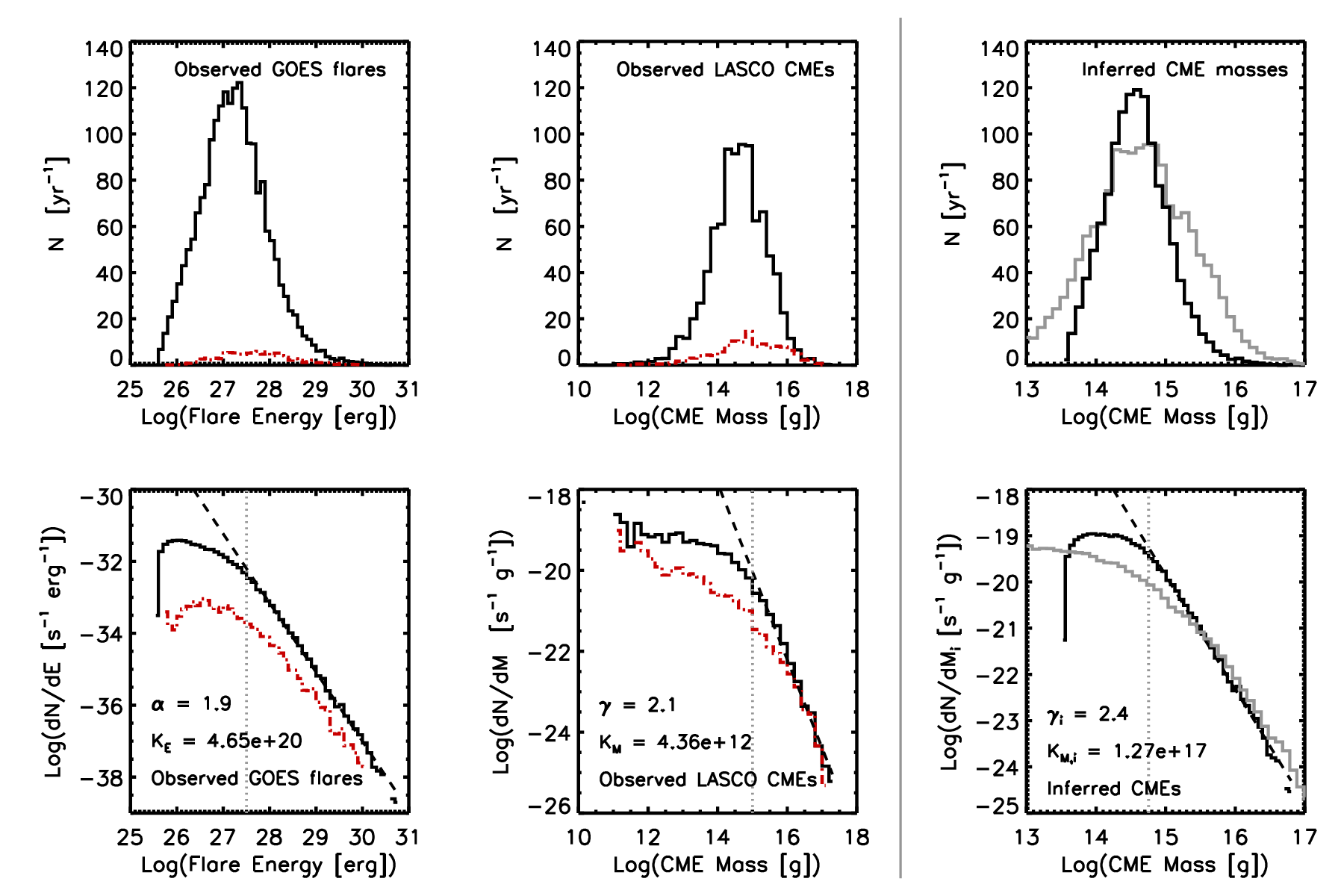}}
\caption{Observed distributions of solar flare energy and of measured solar CME masses, compared to CME masses inferred from the observed 
flares using the methodology described in Section \ref{sec_solarempirical}. In the left column, flare rates as a function of 
energy from the GOES database from 1996-2006 are shown (histogram, upper panel; differential dN/dE distribution, lower 
panel). The middle column shows CME rates as a function of masses from the LASCO database during the same period (histogram, 
upper panel; differential distribution dN/dM, lower panel). In all four of the observed distributions (left and middle columns), black 
curves denote the full distribution, and red dot-dashed curves show the distributions of properties for only those flares and CMEs which are 
associated. The CME mass distribution includes all CMEs in the LASCO database with reported masses plus the Halo CMEs assigned masses by 
\citet{Aarnio:2011a}. Panels in the right column show our inferred CME rate as a function of 
mass, inferred from the observed flare distributions described in Section \ref{sec_solarmassloss}. The black distributions are the inferred 
distributions, while the gray are the observed distributions (as seen in black in the center column), for comparison. In bottom panels, 
vertical dotted lines represent the energy or mass limits above which the distributions are approximately complete.}
\label{fig_solardists}
\end{figure*}

To estimate the solar CME mass-loss rate from flares, we require a flare energy distribution to use with our empirical relation between flare energy 
and CME mass (eq.~\ref{eq_energymass}). In order to do this as we will below for TTSs, we can determine a solar flare frequency f$_{flr}$ as done 
by \citet{Colombo:2007} for the COUP data. The GOES satellite recorded 22,674 solar flares from 1996 to 2006, so the total f$_{flr} =$ 0.065 ksec$^{-1}$ 
(this value is 0.0007 for the flares observed by the COUP; see Sec.~\ref{sec_stellarempirical} below). The energy distribution of observed flare rates 
(number per year as a function of energy) is shown in Figure \ref{fig_solardists}, top left panel. As with the CMEs, we convert this flare energy 
distribution into a differential energy distribution (dN/dE; Fig.~\ref{fig_solardists}, lower left panel) and fit a power law of the form:
\begin{equation}\label{eq_ffreqpowerlaw}
\frac{dN}{dE} = K_{\rm E} E^{-\alpha};
\end{equation}
our fit above an estimated completeness threshold (see below) of $E_{\rm cut}=$10$^{27.5}$ erg (vertical line in Fig.~\ref{fig_solardists}, lower 
left panel), gives $\alpha =$1.92 ($\pm$0.02) and K$_{\rm E} =$ 4.7$\times$10$^{20}$ ($\pm$3\%) in cgs units. 
Our value for $\alpha$ is consistent with those reported in the literature: \citet{Hudson:1991} reports $\alpha =$ 1.8 in $\frac{dn}{dW} = A W^{-\alpha}$, 
where $W$ is the total energy radiated by the flare, and below the GOES detection limit, RHESSI (Ramaty High Energy Solar Spectroscopic Imager) 
data indicate that this distribution extends to lower flare energies and continues to exhibit power-law behavior, with a slope of $\sim$1.5 \citep{Christe:2008}.

Next, we compute the inferred mass distribution of CME rates by using equation \ref{eq_energymass} to convert the energy bins in the flare energy 
distribution (Fig.~\ref{fig_solardists}, left bottom panel) into mass bins, i.e., $dN/dE \rightarrow dN/dM$, where 
\begin{equation}
\frac{dN}{dM} = \frac{dN}{dE} \times \frac{dE}{dM}
\end{equation}
and where $dE/dM$ is given by equation \ref{eq_energymass}. The resulting differential CME mass distribution is shown in Figure \ref{fig_solardists}, 
right bottom panel, and the corresponding non-differential distribution is shown in the right upper panel for ease of comparison with the other 
distributions shown in the top panels. 
The total mass in this empirically determined mass distribution is $\dot{M}_{\rm cme} = \sum dN/dM M = $ 1.558$\times$10$^{18}$g yr$^{-1}$, which 
impressively is equal to the total observed LASCO solar CME mass loss rate (Sec.~\ref{sec_solarmassloss}) to within less than a percent. 

This is reassuring that the CME mass-loss rate inferred directly from observed flares is able to reproduce the directly measured CME mass-loss rate. At 
the same time, as the right column of Figure \ref{fig_solardists} shows, our inferred mass distribution reflects a couple of known biases: first, we 
under-predict the highest mass CMEs because our flare--CME relationship is not taking into account the association probabilities of flares and CMEs 
which increases with greater flare energy. Second, we over-predict the observed occurrence of lower mass CMEs likely because of a bias against the 
detection of the lowest energy flares (in Fig.~\ref{fig_solardists}, dN/dE ``turns over'' at lower energies).

Thus, while our empirical flare-inferred CME mass-loss rate appears to work well at reproducing the directly
measured CME mass-loss rate, we can attempt to account for those flares and CMEs that are missed observationally by 
describing the observed CME rate analytically, as we now discuss.

\subsubsection{Analytically determined solar CME mass-loss rate}\label{sec_solaranalytical}

In an effort to account for the observational biases mentioned above---most importantly the missed low-energy flares and therefore the lowest mass 
CMEs---we can replace the directly observed flare distribution with analytical fits to the observationally complete parts 
of our flare energy and CME mass distributions, and assuming a simple power-law of the form:
\begin{eqnarray}\label{eq_mdot}
\dot{M}_{\rm CME} & = & \int^{M_{\rm max}}_{M_{\rm min}} \frac{dN}{dM_{\rm i}} M_{\rm i} dM \\ 
                  & = & \frac{K_{\rm M,i}}{2-(\gamma_{\rm i})} \left(M_{\rm max}^{2-\gamma_{\rm i}} - M_{\rm min}^{2-\gamma_{\rm i}} \right), 
\end{eqnarray}
where M$_{\rm i}$ represents our flare-inferred CME masses from equation \ref{eq_energymass}, and 
\begin{eqnarray}\label{eq_dndmi}
\frac{dN}{dM_{\rm i}} & = & K_{\rm M,i} M_{\rm i}^{-\gamma_{\rm i}} \\
                      & = & \frac{K_{\rm E}}{\beta K_{\rm M,i}^{(\alpha+\beta-1)/\beta +1}} M_{\rm i}^{(\alpha+\beta-1)/\beta}.
\end{eqnarray}
Equation \ref{eq_mdot} is the integral of the inferred CME mass distribution, i.e., the integrated power law of Figure \ref{fig_solardists}, 
bottom right panel, and $K_E$ and $\beta$ are from equation \ref{eq_energymass} (the fitted values of K$_{\rm M,i}$ and $\gamma_{\rm i}$ are reported 
in the plot legend). For the upper limit of the integral, we take the highest solar CME mass observed, M$_{max}~\sim$6$\times$10$^{16}$ g. To 
determine the value of M$_{min}$, we can either use the observational completeness limit of dN/dE or else simply the minimum observed CME mass.
We begin with the clearest observational constraint, the completeness limit of dN/dE. 
Plugging E$_{\rm cut}$ (10$^{27.5}$ erg) into the flare energy/CME mass relationship (eq.~\ref{eq_energymass}), we find a corresponding 
M$_{\rm min}$ of $\sim$10$^{15}$ g. This represents a very conservative minimum mass; clearly, the observed CME mass distribution contains a 
significant number of CMEs less massive than $\sim$10$^{15}$ g. The resulting mass loss rate, then, as calculated with equation \ref{eq_mdot} is 
6.6$\times$10$^{-16}$ \Msun yr$^{-1}$. Were we to simply set M$_{\rm min}$ equal to the minimum mass CME inferred, $\sim$10$^{13}$ g, we find a 
mass loss rate of 2.2$\times$10$^{-15}$ \Msun yr$^{-1}$, roughly 10\% of the solar wind mass loss rate.

\subsection{T Tauri Star case}\label{sec_stellarmassloss}

In Section \ref{sec_solarmassloss} we demonstrated that the observed distribution of solar flares, when converted to 
CME masses via our empirical relation between flare energy and CME mass (eq.~\ref{eq_energymass}), correctly recovers the observed
distribution of solar CMEs. We moreover developed an analytical representation of the solar flare distribution in an attempt
to account for observational biases in the observed solar flare distribution, and hence in our flare-inferred solar CME mass
distribution. In this section we follow the same procedure, now applied to PMS stellar flares in order to arrive at a PMS
stellar CME mass-loss rate.

\subsubsection{Empirically determined TTS CME mass-loss rate}\label{sec_stellarempirical}

As in the solar case, we can infer an empirical CME mass-loss rate for TTSs by combining our solar-calibrated flare energy/CME mass 
relationship extended into the TTS flare regime (eq.~\ref{eq_energymass} and Fig.~\ref{fig_massenergy}) with an empirical 
flare-energy distribution function for TTS flares. 

\begin{figure*}[ht]
\centerline{\includegraphics[scale=0.5,angle=90]{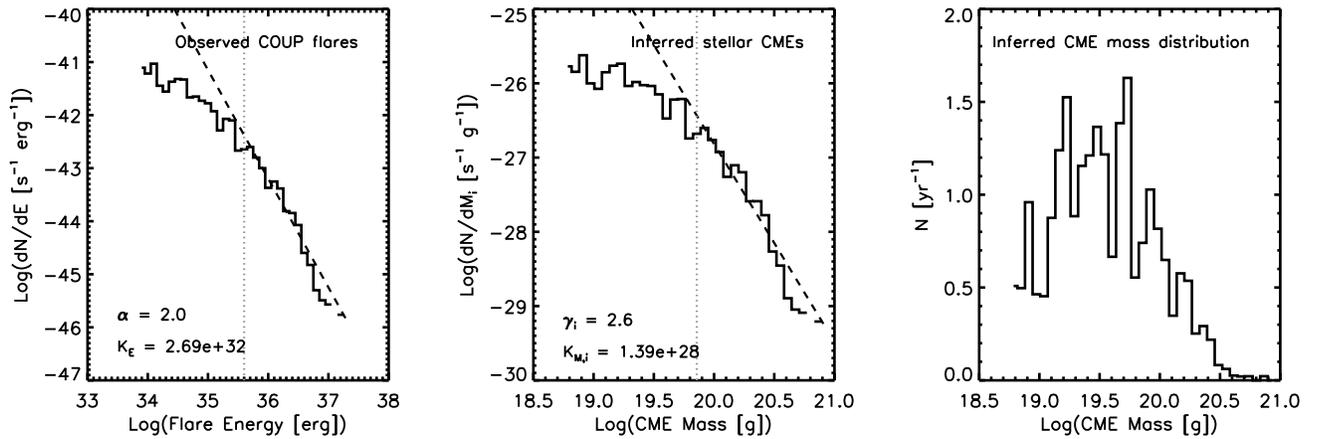}}
\caption{Flare and CME distributions for PMS stellar sample from the COUP. The left panel is the flare frequency distribution of 
\citet{Colombo:2007}, expressed in terms of number of flares of a given energy per time per energy bin ($dN/dE$). In the middle panel, 
we show our inferred TTS CME mass distribution: we have taken the product of dN/dE from the left panel and dE/dM derived 
here in Section \ref{sec_solarecap} to arrive at dN/dM. Then, in the right panel, we show the stellar CME mass distribution 
were it observed over the same time span as the LASCO CMEs (compare to Fig.~\ref{fig_solardists}, upper middle panel).}
\label{fig_stellardists}
\end{figure*}

We utilize the energy distribution of TTS flare rates already measured for the COUP sample by \citet{Colombo:2007}. 
Next, we convert the cumulative distribution of \citet{Colombo:2007} to dN/dE (Fig.~\ref{fig_stellardists}, left panel). Proceeding as in the solar case 
(Sec.~\ref{sec_solarmassloss}), this dN/dE and the CME mass/flare energy relationship (eq.~\ref{eq_energymass} and Fig.~\ref{fig_massenergy}) are used to 
calculate a CME mass distribution (Sec.~\ref{sec_solaranalytical}; Fig.~\ref{fig_stellardists}, middle and right panels). Directly summing this 
inferred CME mass distribution, we obtain a CME mass-loss rate of $\dot{M}_{\rm CME} = \sum dN/dM M =$ 6.2$\times$10$^{-13}$~\Msun yr$^{-1}$. This is an order 
of magnitude more mass-loss than the present-day solar wind. As in the solar case, this likely represents a lower limit on the true TTS CME mass-loss rate due 
to observational biases against the detection of the smallest flares/CMEs. 

\subsubsection{Analytically determined TTS CME mass-loss rates}\label{sec_stellaranalytical}

As in the solar case, the empirical flare-inferred $\dot{M}_{\rm CME}$ presented above is subject to one clearly problematic bias: the X-ray 
flare detection limit. Below this limit, $E_{\rm cut}$, as in the solar case, the flare rate distribution likely continues to increase 
to ever lower flare energies; if this behavior is as on the Sun, the distribution likely follows a power law over many dex in energy. The empirical 
$\dot{M}_{\rm CME}$ above, then, provides a lower limit for the mass loss. Here we present an analytical solution that represents an upper limit 
by attempting to account for the under-detected, lower energy/lower mass flare/CME events.

The power law fit to the COUP flare energy/event rate distribution above the completeness limit of $E_{\rm cut} =$ 10$^{35.6}$ erg is as described 
in equation \ref{eq_ffreqpowerlaw}, where $\alpha =$2.1 \citep{Colombo:2007}. We do not have directly observed bounds on TTS CME masses, 
so we instead perform our distribution integration over flare energy and using our relation between flare energy and CME mass (eq.~\ref{eq_energymass}). 
The mass loss rate is then (equivalent to eq.~\ref{eq_mdot}, with substitutions):
\begin{eqnarray}\label{eq_MdE}
\dot{M}_{\rm CME} & = & \int^{E_{\rm max}}_{E_{\rm min}} \frac{dN}{dE} M_{\rm i} dE \\ 
                  & = & \frac{K_{\rm E}K_{\rm M}}{(\beta-\alpha+1)} \left(E_{\rm max}^{\beta-\alpha+1} - E_{\rm min}^{\beta-\alpha+1} \right). 
\end{eqnarray}

To place a lower limit on the value of E$_{\rm min}$ (and thus an upper limit on $\dot{M}_{\rm CME}$), we find the energy at which the X-ray 
luminosity in flares (L$_{\rm flare}$) is equal to the total stellar X-ray luminosity, $L_X$: 
\begin{equation}\label{eq_lfemin}
L_{\rm flare} = \int^{E_{\rm max}}_{E_{\rm min}} \frac{dN}{dE} EdE = \frac{K_{\rm E}}{2-\alpha} (E_{\rm max}^{2-\alpha} - E_{\rm min}^{2-\alpha}) = L_X.
\end{equation}
For the stellar $L_X$, we take a fiducial value of 2.5$\times$10$^{30}$ erg s$^{-1}$ from the empirical relationship of \citet{Preibisch:2005} for a 
1~M$_\odot$ star. We note that the scatter in L$_{\rm X}$ is rather large \citep[see figure 3 of][]{Preibisch:2005}; the 1$\sigma$ scatter about
their linear regression fit to L$_{\rm X}$ as a function of mass is 0.65 dex. Thus our choice of fiducial $L_X$ could be as low as 5$\times$10$^{29}$ 
erg s$^{-1}$ or as high as 1$\times$10$^{31}$ erg s$^{-1}$. We choose the maximum energy in equation \ref{eq_lfemin} to simply be the most energetic 
flare observed in the COUP sample, $\sim$10$^{37}$ erg. 
Evaluating equation \ref{eq_lfemin}, we find E$_{\rm min} \ga$ 3.4$\times$10$^{29}$ erg. With these parameters, we find an upper limit to the mass 
loss rate for TTS via equation \ref{eq_MdE}: $\dot{M}_{\rm CME}\la$ 3.2$\times$10$^{-10}$ \Msun yr$^{-1}$.

This mass loss rate is most sensitive to the flare frequency/energy distribution slope $\alpha$; were we to vary $\alpha$ within the reported 
uncertainty, the mass loss rates would change by a couple orders of magnitude: for $\alpha$ values of 2.05 and 2.2, $\dot{M}_{\rm CME}$ is 
1.9$\times$10$^{-9}$ \Msun yr$^{-1}$ and 7.1$\times$10$^{-11}$ \Msun yr$^{-1}$, respectively. This is consistent with the solar case and our 
expectation that the CME mass loss rates would not exceed typical wind mass loss rates for young stars.

\subsection{CME mass loss rates compared to steady wind, observed outflow rates}

We have devised a method to infer the CME mass-loss rate from the observed flaring rate together with our empirical relation between flare energy 
and CME mass. Testing this method on the Sun, our empirical method predicts the CME mass loss rate within less than a percent of the observed value. 
We then applied the same method using an analytical power-law representation of the observed flare distribution in order to attempt to account for 
observational biases against the lowest-energy flares (and thus against the lowest mass CMEs). This analytical method 
predicts 2.8 times more mass loss than is observed. With respect to the solar wind mass loss rate \citep[10$^{-14}$ \Msun yr$^{-1}$, ][]{Li:1999}, 
the empirical result (as well as the observed CME mass loss rate) is 4\% of the wind mass loss rate, while the analytical result is that CMEs lose 
mass at 10\% the solar wind rate. Given that the observed CME mass loss rate is itself underpredicted (only half of the CMEs reported in the database 
have measured masses), and the observational bias against the lowest mass flares is present in our empirical result as well as the actual solar data, 
we would anticipate the analytic calculation to be greater than the observed mass lost. As such, these two approaches---empirical and 
analytical---serve as lower and upper limits, respectively.

In the TTS case, our empirically calculated lower limit on the stellar CME mass loss rate is $\sim$6$\times$10$^{-13}$ \Msun yr$^{-1}$. Our analytically 
derived mass loss rates range from $\sim$10$^{-11}-$10$^{-9}$ \Msun yr$^{-1}$ (when taking into account the error on the stellar flare frequency power 
law slope). For comparison, TTS wind mass loss rates have been derived from various line diagnostics: for TW Hya, \citet{Dupree:2005} report 
$\dot{M}_{\rm O~VII} / \phi =$ 2.3$\times$10$^{-11}$ \Msun yr$^{-1}$ and
$\dot{M}_{\rm C~III} / \phi =$ 1.3$\times$10$^{-12}$ \Msun yr$^{-1}$, 
where $\phi$ is the fractional stellar surface area from which the wind originates (e.g., a more collimated, polar wind will have $\phi \lesssim$0.3). 
These outflow rates are less than the H$\alpha$ emission derived accretion rate for TW Hya of 4$\times$10$^{-10}$ \Msun yr$^{-1}$ \citep{Muzerolle:2000}. 
In cases of other accreting systems, wind mass loss rates inferred from accretion signatures in spectral lines range from 10$^{-9}$ to 10$^{-7}$
\citep{Hartigan:1995}.


\section{Angular momentum loss via CMEs}\label{sec_angmom}

Having calculated mass loss rates from PMS stellar CMEs, we now focus on whether or how these episodic mass loss events could impact the angular 
momentum evolution of a TTS. We have determined order-of-magnitude CME mass-loss rates, calculated empirically as well as analytically, giving 
approximate lower and upper limits of $\sim$10$^{-12}$ and $\sim$10$^{-9}$ \Msun yr$^{-1}$, respectively (Sections \ref{sec_stellarempirical} and 
\ref{sec_stellaranalytical}). In this section, we estimate magnetic lever arm lengths to calculate torques and assess their relative importance in spin 
evolution.

\subsection{What is the magnetic lever arm length?}

The torque on the star, due to a steady, magnetized wind can be written
\begin{eqnarray}
\label{eq_tauw}
T_{\rm w} = - \dot M_{\rm w} \Omega_* r_{\rm A}^2,
\end{eqnarray}
where $\dot M_{\rm w}$ is the mass loss rate in a steady wind, $\Omega_*$ is the angular rotation rate of the star, and $r_{\rm A}$ is the
average ``lever arm'' radius in the wind \citep[e.g.,][and references therein]{Matt:2008b}.

Computing the lever arm radius precisely requires knowledge of the three-dimensional 
flow structure and magnetic field configuration \citep[e.g.,][]{Mestel:1984}. For 
time-variable flows, such as CMEs, the situation is even more complicated.  Thus, here 
we will only estimate the lever arm radius that is 
appropriate for an ensemble of CMEs and that is based upon our current understanding of 
magnetized stellar winds.  Based on simulations of 
steady-state, ideal MHD winds from stars with dipole magnetic field and rotating at 10\% of 
breakup speed 
\citep{Matt:2008b,Matt:2012} found
\begin{eqnarray}\label{eq_ra1}
{r_{\rm A} \over R_*} \approx 2.1 \left(\eta {B_*^2 R_*^2 \over \dot M_{\rm
   CME} v_{\rm esc}}\right)^{0.22},
\end{eqnarray}
where $B_*$ is the equatorial field strength of the global/dipolar magnetic field at the stellar surface, $R_*$ is the radius of the star, 
and $v_{\rm esc}$ is the escape speed from the stellar surface.  Here, we have introduced a factor $\eta$, in order to account for the difference 
in the effective lever arm length for an ensemble of eruptive mass loss events ($\dot M_{\rm CME}$), as compared to a steady wind ($\dot M_{\rm w}$).

There are several factors that modify the effective lever arm length in an eruptive outflow, compared to a steady one.  The first is that the mass 
loss is not continuous, but happens in discrete bursts/ejections. For simplicity, we consider that the mass loss occurs with an average rate of 
$\dot M_{\rm w} = \dot M_{\rm CME}$, but occuring via discrete bursts that happen for a fraction $f_t$ of the time ($f_t < 1$).  Thus, each burst 
has an instantaneous mass loss rate of $f_t^{-1} \dot M_{\rm CME}$, and the factor of $\eta$ in equation \ref{eq_ra1} should include a factor of 
$f_t$, in order to take this into account.  It is clear that a time-dependent wind is less efficient at extracting angular momentum than a steady 
wind with the same average mass loss rate.  In the inferred distribution of CMEs (see Fig.~\ref{fig_stellardists}, middle panel), 
the mass loss is dominated by the lowest energy events, of which there are $\sim 10$ per year.  To estimate $f_t$, we also need to specify the 
duration of each event---i.e., the duration over which we can think of the CME as a steady wind from the surface of the star.  As an order of 
magnitude estimate, we consider the time for a CME traveling several hundred km/s to cross several stellar radii, which is of the order of one hour.  
For 10 events per year with a duration of one hour, $f_t \sim 10^{-3}$.  To calculate our upper limit to $\dot M_{\rm CME}$ in 
Section \ref{sec_stellaranalytical}, we considered that the flare rate distribution extends to much lower energies than observed.  In this case, the 
event frequencies of the lowest energy flares are much higher, implying a larger value for $f_t$.  

The second factor modifying the effective lever arm length in an eruptive flow is that the CME events are not globally distributed in the corona, 
but take place over a fraction of the total solid angle of whole sky.  More importantly for the angular momentum flow, not all of the stellar magnetic 
flux participates in the azimuthal acceleration of the outflow. To take this into account, the factor $\eta$ in equation \ref{eq_ra1} should include 
a factor of $f_\psi^2$, where $f_\psi$ is the fraction of (unsigned) magnetic flux participating in the CME flow, compared to the total stellar surface 
magnetic flux associated with the global/dipole field.  The unsigned flux participating in a CME should approximately equal the geometric area times the 
field strength of the underlying active region.  Generally speaking, and almost by definition, the magnetic field strengths of active regions are much 
stronger than the global field strength, while the areas are much smaller than the total stellar surface area.  The product of these two approximates 
$f_\psi$, and it is not clear in general whether this will be very different from unity.  For the purpose of this section, we assume $f_\psi \sim 0.1$--1.0.

Other factors that influence the effective lever arm length is the angle that the direction of the CME flow makes with respect to the rotation axis of 
the star and that the acceleration of the CMEs may be significantly different than the acceleration of the steady wind in the simulations of Matt et al.
Given the much larger uncertainties discussed above, we neglect these effects for the present analysis and approximate $\eta=f_tf_\psi^2$.  We consider 
here a range of $10^{-3}\le \eta \le 1$ but acknowledge that an even wider range may be possible.  Note that the case of $\eta = 1$ is equivalent to the 
case of a steady wind.

To compute the lever arm length as a function of time, we use a model for the evolution of a 1 $M_\odot$ star from \citet{Siess:2000}, which gives the 
stellar radius $R_*$ (and thus surface escape speed $v_{\rm esc}$) as a function of stellar age.  The upper left panel of Figure \ref{fig_CMEtorque_m1.0} 
shows the evolution of $R_*$.  To compute the lever arm length, we must specify the surface magnetic field strength, mass loss rate, and $\eta$.  For 
simplicity, we adopt a singular value of $B_* = 600$ G \citep[corresponding to the equatorial field strength of the dipole field measured on BP Tau;][]{Donati:2008}, 
and assume that it is constant in time.  The range of possible values of $B_*$ for TTSs, as well as how this may evolve in time, is not well constrained 
observationally, which introduces an uncertainty in our calculated values.  The upper right panel of Figure \ref{fig_CMEtorque_m1.0} shows the Alfv\'en radius as 
a function of time, for four combinations of the extreme values of mass loss rate and $\eta$ discussed above.  In the Figure, the vertical dash-triple-dotted 
lines mark the approximate age range of TTSs (for which this calculation is valid), corresponding roughly to an age of a few Myr, plus or minus a few 
Myr.  Since the mass loss rates we are considering correspond to TTSs, the calculation is only valid/relevant between these two vertical lines, but we 
show a wider range of ages for context/illustrative purposes.  It is clear that the range of $\eta$ and $\dot M_{\rm CME}$ considered corresponds to a range of a 
factor of about 13 in the possible Alfv\'en radii at any given age.  In the age range of TTSs, the Alfv\'en radius could be in the range of 3--70 $R_*$, 
depending on the wind parameters.

Observationally, the largest magnetic structures are seen to exist at distance scales comparable to Alfv\'en radii calculated here: 
the magnetic loops of \citet{Favata:2005} were inferred to be from 0.1--55 $R_*$ in extent (assuming these were indeed single loops), and cool prominences have 
been seen to extend up to a few stellar radii from TTS \citep{Massi:2008,Skelly:2008}. At later evolutionary stages, solar type stars are still seen to have extended, 
suspended material: the 50 Myr old \citep{Luhman:2005} solar analog AB Dor was observed to have cool, corotating clouds at distances from $\sim$9-20 R$_{*}$ from the 
rotation axis, and were estimated to have masses $\gtrsim$10$^{18}$g \citep{Cameron:1989}. \citet{Dunstone:2006} observed prominences on Speedy Mic, a 30 Myr old K3 
dwarf, and estimated masses to be $\sim$2$\times$10$^{17}$g. In stellar outflows, the Alfv\'en radii are typically comparable to or slightly larger than the size of 
the largest closed magnetic loops \citep[e.g.,][]{Mestel:1987}. Thus, the measured sizes of post-flare loops and prominences may represent an approximate lower limit 
on the Alfv\'en radii for these systems.

\subsection{How important is the torque?}

The spin rate of stars will change in response to external torques and changes in the structure of the star (and subsequent change in the stellar moment of inertia).
The conservation of angular momentum for an isolated star rotating as a solid body can be expressed as
\begin{eqnarray}
\label{eq_pseudo}
I_* {d \Omega_* \over d t} = T_{\rm w} - \Omega_* {d I_* \over d t},
\end{eqnarray} 
where $I_* \equiv k^2 M_* R_*^2$ is the moment of inertia of the star, $\Omega_*$ is the angular spin rate of the star, $T_{\rm w}$ is the wind torque 
(eq.~\ref{eq_tauw}); here, we're treating the CME outflow like a wind, so $T_{\rm w} = T_{\rm CME}$), and $k^2$ is the normalized radius of gyration 
(defined by the radial mass distribution).  For a star in a tight binary system or a star that is still actively accreting from a disk, there will be additional 
torque terms in equation \ref{eq_pseudo}, which we neglect here in order to isolate the effects of CMEs alone.

To compute the relative strength of the terms in this equation, we use the 1 $M_\odot$ \citet{Siess:2000} model to specify the evolution of $k^2$ and $R_*$.  In this 
and in most other PMS stellar evolution models, the star is fully or nearly fully convective during the first $\sim$10 Myr of evolution.  In convective regions, the mixing 
of material redistributes angular momentum on a timescale that is much shorter than evolutionary timescales.  Thus, external torques effectively act upon the entire mass of 
the convection zone; for a fully convective PMS star, the external torques act on the entire star (as opposed to acting only on a thin shell that is rotationally decoupled 
from the interior). There have been observations of possible differential rotation in a few TTSs, with amplitudes of $\sim 10$\% of the bulk/average rotation rate 
\citep[e.g.,][]{Herbst:2006}, but in the vast majority of TTSs studied no measureable differential rotation is observed. Thus here we follow convention and treat 
the star as a solid-body rotator, where $\Omega_*$ represents the bulk rotation rate, and the error associated with this approximation is much smaller than for other 
unknowns (such as the value of $\eta$).  In order to specify $\Omega_*$, we simply assume that the star is always rotating at a fraction of $f=10$\% of breakup speed, in order 
to approximately represent the ``slow rotators'' in observed TTS spin distributions \citep[e.g.,][]{herbst07}.  This means that the rotation period is assumed to be
\begin{eqnarray}
P_* = {2 \pi \over \Omega_*} \approx 1.17 \; {\rm days} \; \left({0.1\over f}\right)
    \left({R_*\over R_\odot}\right)^{3/2} \left({M_\odot \over M_*}\right)^{1/2}
\end{eqnarray}
Since both terms on the right hand side of equation \ref{eq_pseudo} are proportional to $\Omega_*$, the relative value of these two terms is independent on the 
assumed value of $\Omega_*$.

\begin{figure*}[ht]
\centerline{\includegraphics[width=0.9\textwidth]{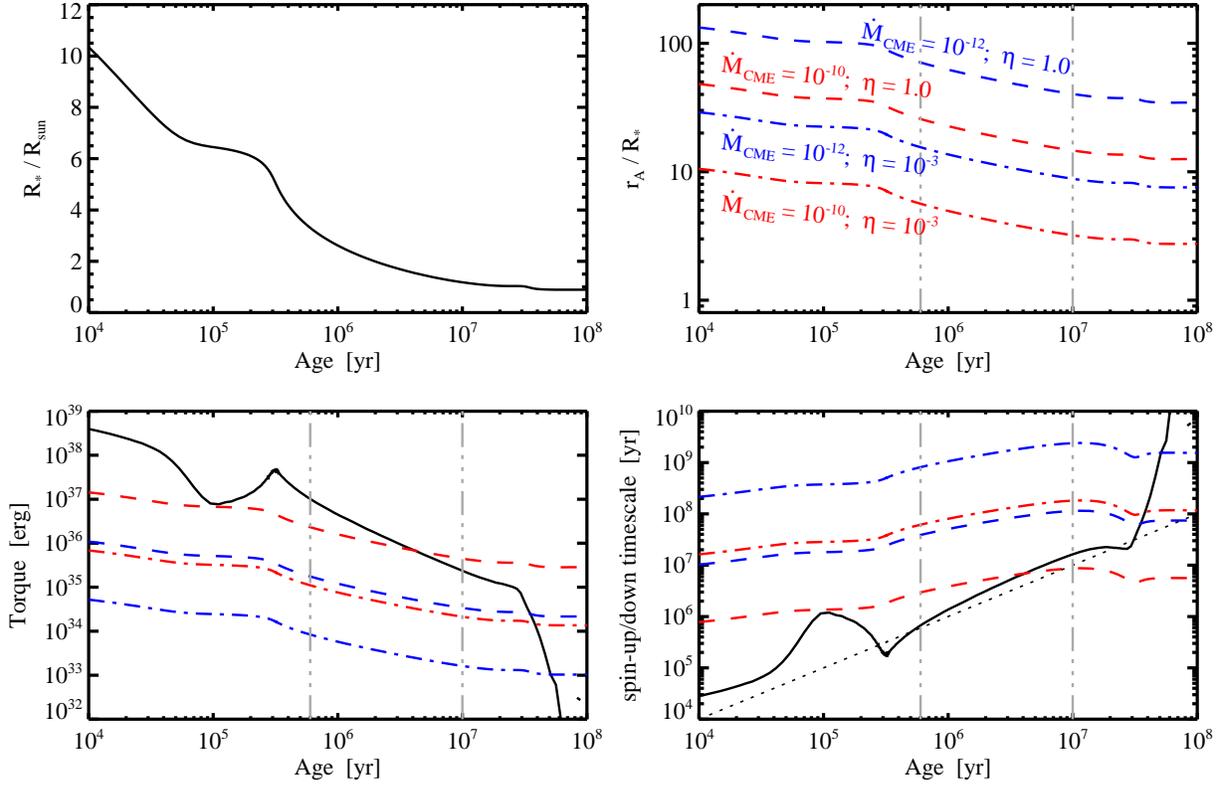}}
\caption{Upper left panel: Evolution of the photospheric radius of a one solar mass star, as a function of age, from the model 
tracks of \citet{Siess:2000}.  Upper right panel:  Effective magnetic lever arm (Alfv\'en) radius in CME outflows from the 
same star, with a magnetic field strength of 600 G and four different combinations of mass loss rate ($\dot M_{\rm CME}$) and 
efficiency factor ($\eta$), as indicated on the plot ($\dot M_{\rm CME}$ is given in units of $M_\odot$ yr$^{-1}$).  The vertical 
dash-triple-dotted lines show the approximate age range of the T Tauri phase.  Lower left panel:  The solid line shows the 
"pseudo-torque," which shows the contribution to the evolution of stellar rotation rate that is due to the contraction and 
changes of internal structure of the star(see text).  The colored lines show the CME-outflow torque, which acts to spin down 
the star, corresponding to different outflow parameters (line styles correspond to the upper right panel).  Lower right panel:  
The solid line shows the spin-up timescale, due only to the contraction of changes of internal structure of the star, as a 
function of stellar age.  The colored lines show the spin-down timescale due to CME outflows, corresponding to different 
outflow parameters (corresponding to the upper right panel).  The dashed line corresponds to a spin-timescale equal to the age 
of the star.}
\label{fig_CMEtorque_m1.0}
\end{figure*}

The bottom left panel of Figure \ref{fig_CMEtorque_m1.0} compares the two terms on the right hand side of equation \ref{eq_pseudo} as a function of time.  The broken 
lines show the value of $-T_{\rm CME}$, where we have multiplied by $-1$ in order to compare the absolute value of the terms on a logarithmic scale.  The different line 
styles and colors correspond to the assumptions and range of possible values of stellar wind parameters discussed in the previous section and labeled in the top right 
panel of the Figure.  The stellar wind torque always acts to spin down the star (decrease $\Omega_*$).  The solid line in the bottom left panel shows the last term on 
the right-hand side of equation \ref{eq_pseudo}.  This term has the dimensions of torque and can be thought of as a pseudo torque, which describes how the spin rate 
of the star changes due to changes in the moment of inertia (i.e., even if total angular momentum were conserved), and which acts to spin up the star (increase $\Omega_*$).

In order to assess how the terms in equation \ref{eq_pseudo} may affect the stellar spin rate, it is useful to compute a spin-up or spin-down timescale by dividing 
the total angular momentum by the torque.  By rearranging equation \ref{eq_pseudo}, the spin-up/down time can be written
\begin{eqnarray}
\label{eq_spinupdown}
{\Omega_* \over \dot \Omega_*} = {\Omega_* \over T_{\rm CME}/I_* - 
                                \Omega_*/I_* (d I_* / d t)}.
\end{eqnarray}
When the CME-outflow torque dominates, the star will spin down on a timescale
\begin{eqnarray} 
\label{eq_spindown}
\tau_{\rm CME} = {I_* \Omega_* \over T_{\rm CME}} 
                      = k^2 \left({M_* \over \dot M_{\rm CME}}\right) \left({R_* \over r_{\rm A}}\right)^2. \\
\end{eqnarray}
By contrast, when the changes in stellar structure dominate, the star will spin up on a timescale
\begin{eqnarray}
\label{eq_spinup}
\tau_{\rm struct} = - {I_* \over(d I_* / d t)}.
\end{eqnarray}
One advantage of looking at things in this way is that it is independent of the current spin rate of the star.  In the bottom right 
panel of Figure \ref{fig_CMEtorque_m1.0}, we show $\tau_{\rm struct}$ (solid line) and $\tau_{\rm CME}$ (broken lines, corresponding 
to the various wind parameters indicated in the upper right panel), as a function of stellar age, for the 1 $M_\odot$ Siess et al.\ model.  
At any given time, the shortest timescale is the dominant one, and if it is comparable to or shorter than the age of the star (shown 
as a dotted line), it will have a noticible effect on the stellar spin rate.

It is clear from Figure \ref{fig_CMEtorque_m1.0} that the stellar contraction is expected to 
effectively spin the star up during the first 30 Myr (in the absence of any external torques).
The Figure shows that this pseudo-torque from stellar contraction completely dominates over the 
angular momentum loss associated with CMEs, for most of the range of parameters considered here.  
The only exception is that, for the highest mass outflow rates allowable by our analysis, and for 
$\eta \sim 1$, the angular momentum loss by CMEs becomes comparable to the contraction pseudo-torque 
at an age of a few million years and dominates it thereafter.  

Many of the stars in the COUP sample are actively accreting from a surrounding accretion disk.  It is therefore interesting to consider the 
angular momentum exchange between the star and disk, arising from this interaction.  \citet{Matt:2008c} demonstrated that the spin-down torque 
associated with the magnetic connection between the star and disk will be less than the torque from a stellar outflow, in the case where the 
Alfv\'en radius is not too large and when the magnetic coupling to the disk is strong.  Using their ``preferred'' values for the coupling, 
they demonstrated that a stellar outflow will carry away more angular momentum than the star-disk interaction, as long as $r_{\rm A} < 84 R_*$.  
As demonstrated in the upper right panel of Figure \ref{fig_CMEtorque_m1.0}, $r_{\rm A} \le 70 R_*$ for all of the outflow parameters considered 
here, so the spin-down torque due to a star-disk magnetic connection should be negligible\footnote{The analysis of \citet{Matt:2008c} is valid 
only for $\eta = 1$.  It could be modified to consider $\eta \ne 1$, but the overall results are not affected.}.  Accreting stars can also 
experience strong spin-up torques, mainly associated with the accretion of high specific angular momentum from nearly-Keplerian disks.  The 
spin-up torque from accretion is given by $\dot M_{\rm a} \sqrt{G M_* R_{\rm t}}$, where $\dot M_{\rm a}$ is the accretion rate, and $R_{\rm t}$ 
is the radial location of the inner truncation of disk \citep[e.g.,][]{Ghosh:1978, Matt:2005b}.  As an example, for the star considered here, 
accreting at a rate of $10^{-8} M_\odot$yr$^{-1}$, with a truncation radius of a few stellar radii, and at an age of $6 \times 10^5$ years 
(values appropriate for the early TTS phase), the resulting accretion torque is $\sim 10^{37}$ erg.  This spin-up toruqe is comparable to the 
contraction pseudo-torque at the same age (lower left panel of Figure \ref{fig_CMEtorque_m1.0}).  

Thus, depending on the stellar age, accretion rate, etc., the processes acting to spin up the star may be dominated by either contraction or 
accretion.  In order to counteract this spin-up torque and to explain the observations of slow rotators at all TTS ages, the mass outflow rates 
must be high, the magnetic field strengths must be large, and/or additional spin-down torques (that are not considered here) must be present.  
The CME outflows inferred in the present work could be important near the end of the TTS phase, in the case of weakly- or non-accreting stars, 
and if the mass loss rate is near our upper limit ($\dot M_{\rm CME} \ga 10^{-10} M_\odot$ yr$^{-1}$) and $\eta \sim 1$.


\section{Discussion}\label{sec_discus}

In this work, we derive mass-loss rates for CMEs from solar type PMS stars. This is a first attempt to estimate PMS mass-loss via this mechanism 
and to calibrate it explicitly against the Sun and against observations of flaring magnetic loops on PMS stars. We find the total CME mass-loss rate 
to be in the range of $10^{-12}$ to $10^{-9}$ \Msun yr$^{-1}$. This is relatively modest by comparison with mass loss rates observed for TTS winds. 
However, the magnetic lever arms associated with the largest PMS CMEs, as inferred from the COUP sample stars and scaled from the Sun, are large---up 
to tens of stellar radii---so the resulting torques can be significant. 
Assuming the CME mass loss rate for a typical $\sim$1~\Msun\ TTS is $\lesssim10^{-10}$ \Msun yr$^{-1}$, the associated spin-down torque is likely too 
small to counteract the spin-up effects of contraction and accretion in the T Tauri phase. However, the spin-down torque from CMEs could be important 
after an age of a few Myr if the mass loss rate is $\gtrsim10^{-10}$ \Msun yr$^{-1}$, the accretion rate is low, and the angular momentum loss is 
efficient. Our estimates of the angular momentum loss efficiency (i.e., the magnetic lever arm radii; see Section \ref{sec_angmom}) are based upon 
axisymmetric, steady-state wind calculations, and 3-dimensional simulations are needed to improve these estimates.

Our stellar CME model is derived from relating solar activity to stellar; in one sense, we immediately underestimate mass losses via CMEs by only 
selecting solar CMEs that occurred with flares. We find, however, that much of the mass loss comes from the CMEs associated with flares. The most 
massive CMEs are associated with the most powerful flares, and the most energetic flares are the most often associated with CMEs \citep{Aarnio:2011a}. 
Thus, using stellar flare activity to determine a stellar CME rate likely represents a lower bound on the stellar CME activity. 

One of the most pivotal parts of this calculation is the extrapolation of the CME mass/flare energy relationship. The adopted direct extrapolation, 
while crossing many orders of magnitude, is indeed physically motivated, and introduces the fewest new assumptions. We are able to, using the COUP 
``superflare'' observations, justify the extrapolation up to TTS flare energies; our premise is that these loops represent proxies for stellar 
CME masses as the loops exist within the density and height regimes from which CMEs are generally launched on the Sun. The consistency with the solar 
CME mass/flare energy relationship is particularly compelling given that it spans six dex in flare energy.

In adopting the stellar X-ray flare frequency relationship of \citep{Colombo:2007}, we have generated an ``ensemble mass loss rate''; in doing this, we 
have neglegted to separate out any mass-dependent characteristics of the distribution. There would be, however, significant differences in the flare 
activity for stars of the same mass with different rotation rates. Given the mixed sample of classical and weak-line TTS in the COUP, there are also 
self-absorption effects \citep[by the accretion columns, ][ or a circumstellar disk inclined to our line of sight]{Gregory:2007} which could introduce 
bias into a flare frequency distribution. It was our intent in using the COUP sample to diminish all of these observational effects via the large number 
of stars observed in various activity and evolutionary phases, as well as inclinations and rotation rates. Future work could include analyzing subsets 
of the COUP dataset to see whether there is an X-ray luminosity dependence on the flare rate (and thus, a difference in CME mass loss rates as well).

As noted in Section \ref{sec_stellaranalytical}, one particular difficulty with this analysis was the sensitivity of our analytical solutions to the 
value of the power law slope. Within the errors on these quantities, we find our mass loss rates derived analytically could span two orders of magnitude. 
The empirically derived mass loss rates, while derived from incomplete samples, represent robust lower limits. In the solar case, our analytical mass 
loss rate did predict more mass loss than was observed in the LASCO database, but as discussed in Section \ref{sec_solarecap}, the LASCO database itself 
is missing many CME masses, so the total mass loss rate is undoubtedly higher. Despite this apparent overprediction, we still derive a mass loss rate 
below the steady solar wind mass loss rate, which is as anticipated. 

Observing stellar CMEs would be ideal for addressing the most uncertain parameters in this work, but presently, understanding the observational signatures 
and being able to detect a stellar CME remain outstanding problems. Efforts have been made with EUV spectroscopy; \citet{Leitzinger:2011} point out a lack 
of simultaneous solar spectra during flare/CME events, creating great difficulty for the interpretation of stellar data. Additionally, managing to catch a 
stellar CME with the right set of parameters for observation (e.g., density and projection) requires sufficient observing time to increase the probability 
of detection. For a steady, hot (1 MK), coronal wind, \citet{Matt:2007} calculate a mass loss rate of 1$\times$10$^{-9}$ \Msun yr$^{-1}$ would produce 
strong EUV and X-ray emission, much higher than is observed. Our highest mass loss rate estimate, $\sim$10$^{-9}$ \Msun yr$^{-1}$, is probably unlikely 
given the strong observational signature that would attend such a mass loss rate. Interestingly, however, the authors also found that decreasing the mass 
loss rate by a factor of 10 decreased the excess emission by a factor of 100.  Given the ``bursty'' nature of our CMEs, and the likely high cadence of these 
bursts, and potentially cooler temperatures (solar CMEs are observed to rapidly expand as they travel away from the Sun), a significant level of mass loss 
could easily escape detection.

The COUP ``superflaring'' targets were generally weak lined TTS \citep{Aarnio:2010}; it is likely that accretors could show an enhanced activity rate and 
thus more frequent CMEs, as well as accretion-driven winds \citep{Matt:2008b} which could further deplete angular momentum. The presence of a close-in disk 
in accreting systems might mitigate the enhanced CME rate with reduced lever arm lengths, however modeling work \citep{Orlando:2011} has shown star-disk 
flaring can disrupt circumstellar disk material, causing an MRI instability that then generates an accretion flow. Thus, in a cyclical fashion, large scale 
flaring can lead to accretion, the accretion then powering winds, outflows, and more reconnection, resulting in further mass/angular momentum loss. 
\citet{Cranmer:2009} illustrate one mechanism by which accretion could power a wind: infalling material impacting the stellar surface would generate 
Alfv{\'e}n waves, the propagation of which could trigger reconnection, accretion thus powering activity.


\section{Summary and conclusions}\label{sec_conclusion}

This represents a preliminary effort at calculating mass and angular momentum losses for T Tauri stars via scaled-up, solar-analog coronal mass ejections. 
Our calculations are based on well-observed flares among a large sample of solar-type pre-main-sequence stars, and our analysis methods have been tested for 
the most well-understood case, the Sun. Beginning with an established relationship between solar flare energies and CME masses \citep{Aarnio:2011a}, we use 
the observed flare frequency distribution to infer a CME mass distribution. We are able to infer empirically and analytically solar CME mass loss rates that 
are consistent with observations, and also consistent with expectations for a mass loss rate once observational biases are carefully accounted for. 

Our lower limit on the mass shed via CMEs in a generic TTS case is 6.2$\times$10$^{-13}$ \Msun yr$^{-1}$. To obtain an upper limit on the stellar CME mass 
loss rate, we calculate analytic solutions to account for observational biases (most importantly, the stellar X-ray flare detection limit); our upper limit to 
the mass loss rate via CMEs is $\dot{M}_{\rm CME} \lesssim 2 \times 10^{-9}$ \Msun yr$^{-1}$.

Finally, we assess the resulting torque against the star's rotation provided by the CME mass loss. We find that within the ranges of mass loss rates and 
effective lever arm lengths (reflecting the difference in steady versus ``bursty'' mass loss) being considered, the resulting Alfv\'en radii would span a 
large range, 3--70 R$_{*}$, consistent with previous inferences of the physical sizes of large-scale magnetic structures. We find that near our upper limit 
mass loss case ($\gtrsim$10$^{-10}$ \Msun yr$^{-1}$), this mechanism could be effective for slowing stellar rotation on a timescale comparable to TTS
lifetimes. 

\acknowledgments
We acknowledge NSF grant AST-0808072 (K.~Stassun, PI).


\end{document}